\documentclass[prx,amsmath,amssymb,superscriptaddress,twocolumn]{revtex4-1}
\usepackage{graphics}
\usepackage{graphicx}
\usepackage{bm}
\usepackage{amsmath}
\usepackage{amsfonts,amssymb,amsmath}        
\usepackage{textcomp}
\usepackage{gensymb}
\usepackage{epstopdf}
\usepackage{color}
\usepackage{ulem}

\newcommand{\be}{\begin{equation}}
\newcommand{\ee}{\end{equation}}
\newcommand{\bq}{\begin{eqnarray}}
\newcommand{\eq}{\end{eqnarray}}

\newcommand{\rf}[1]{(\ref{#1})}

\DeclareMathAlphabet{\mathpzc}{OT1}{pzc}{m}{it} 
\begin{document}
\title{Self-organized pseudo-graphene on grain boundaries in topological band insulators}
\author{Robert-Jan Slager}
\affiliation{Instituut-Lorentz for Theoretical Physics, Universiteit Leiden,
P.O. Box 9506, 2300 RA Leiden, The Netherlands}
\author{Vladimir Juri\v ci\' c}
\affiliation{Nordita,  Center for Quantum Materials,  KTH Royal Institute of Technology and Stockholm University,
Roslagstullsbacken 23,  10691 Stockholm,  Sweden}
\author{Ville Lahtinen}
\affiliation{Institute for Theoretical Physics, University of Amsterdam,
Science Park 904, 1090 GL Amsterdam, The Netherlands}
\affiliation{Dahlem Center for Complex Quantum Systems, Freie Universit\"{a}t Berlin,
14195 Berlin, Germany }
\author{Jan Zaanen}
\affiliation{Instituut-Lorentz for Theoretical Physics, Universiteit Leiden,
P.O. Box 9506, 2300 RA Leiden, The Netherlands}

\begin{abstract}
Semi-metals are characterized by nodal band structures that give rise to exotic electronic properties. The stability of Dirac semi-metals, such as graphene in two spatial dimensions (2D), requires the presence of lattice symmetries, while akin to the surface states of topological band insulators, Weyl semi-metals in three spatial dimensions (3D) are protected by band topology. Here we show that in the bulk of topological band insulators, self-organized topologically protected semi-metals can emerge along a grain boundary, a ubiquitous extended lattice defect in any crystalline material. In addition to experimentally accessible electronic transport measurements, these states exhibit valley anomaly in 2D  influencing edge spin transport, whereas in 3D they appear as graphene-like states that may exhibit an odd-integer quantum Hall effect. The general mechanism underlying these novel semi-metals -- the hybridization of spinon modes bound to the grain boundary -- suggests that topological semi-metals can emerge in any topological material where lattice dislocations bind localized topological modes.
\end{abstract} 

\maketitle

Graphene \cite{graphene-review} and topological band insulators (TBIs) \cite{Hasan10,Qi11} show exotic electronic transport properties that have motivated the search for other materials exhibiting similar semi-metallic features. Semi-metals are described by electronic band structures where the bands touch at isolated points or lines in the Brillouin zone (BZ). In graphene, a 2D honeycomb lattice of carbon atoms, or in Dirac semi-metals in 3D, the bulk hosts a pair of pseudorelativistic gapless Dirac fermions, while the surface states of TBIs feature in general gapless Weyl fermions -- chiral massless particles extensively studied in high-energy physics for the description of neutrinos. Recently, the latter have been discovered also in the bulk of 3D materials known as Weyl semi-metals \cite{Xu15,Lv15,Burkov15, Burkov11}. While the energy spectra in all cases resemble each other, the stability of their band structures has a dramatically different origin. In Dirac semi-metals the stability of the Fermi surface relies on lattice symmetries, while Weyl semi-metals and surface states of TBIs are protected by the topology of the bulk band structure. Therefore, it is of both fundamental and practical importance to answer the following question: Can {\it topologically} protected Weyl fermions also appear in the bulk of lower dimensional systems?

We here provide an affirmative answer to this question by showing that grain boundaries (GB) -- ubiquitous crystal defects in real materials that are usually considered as detrimental for their properties -- can host time-reversal symmetry (TRS) protected  topological semi-metals. GBs arise at the interface of two crystal regions (grains) whose lattice vectors are misaligned by an angle $\theta$, as illustrated in Fig.\ 1A. For small opening angles a GB can be viewed as lattice dislocations described by Burgers vector $\mathbf{b}$ arranged on an array of spacing $d=|\mathbf{b}|/(\tan \theta)$. While GBs are usually considered as unwanted disorder, they have also been used experimentally as probes of the superconducting state in high-temperature superconductors \cite{mannhart,kirtley, hilgenkamp}. Recently, they have also been suggested for engineering thermoelectric devices \cite{Science-April2015} and for experimentally tuning the surface states in a 3D TBI \cite{Liu14Y}. Our main result is that extended GB lattice defects can host stable self-organized states of matter. We show that GBs in the bulk of 2D and 3D TBIs  can realize stable TRS protected 1D and 2D semi-metals, respectively, which, in contrast to Dirac fermions in graphene, do not exhibit pseudospin degeneracy. These "halved" graphene-like states can be experimentally observed by measuring their characteristic conductance through otherwise insulating bulk. They are also intimately connected to the TBI surface states and can influence surface transport: 1D GBs in the bulk of a 2D TBIs exhibit a valley anomaly that under an applied electric field results in a helical imbalance of the TBI edge states on the two grains connected by the GB. Furthermore, 2D GBs in the bulk of 3D TBIs may exhibit an odd-integer quantum Hall effect.

\begin{figure*}
\includegraphics[width=\textwidth]{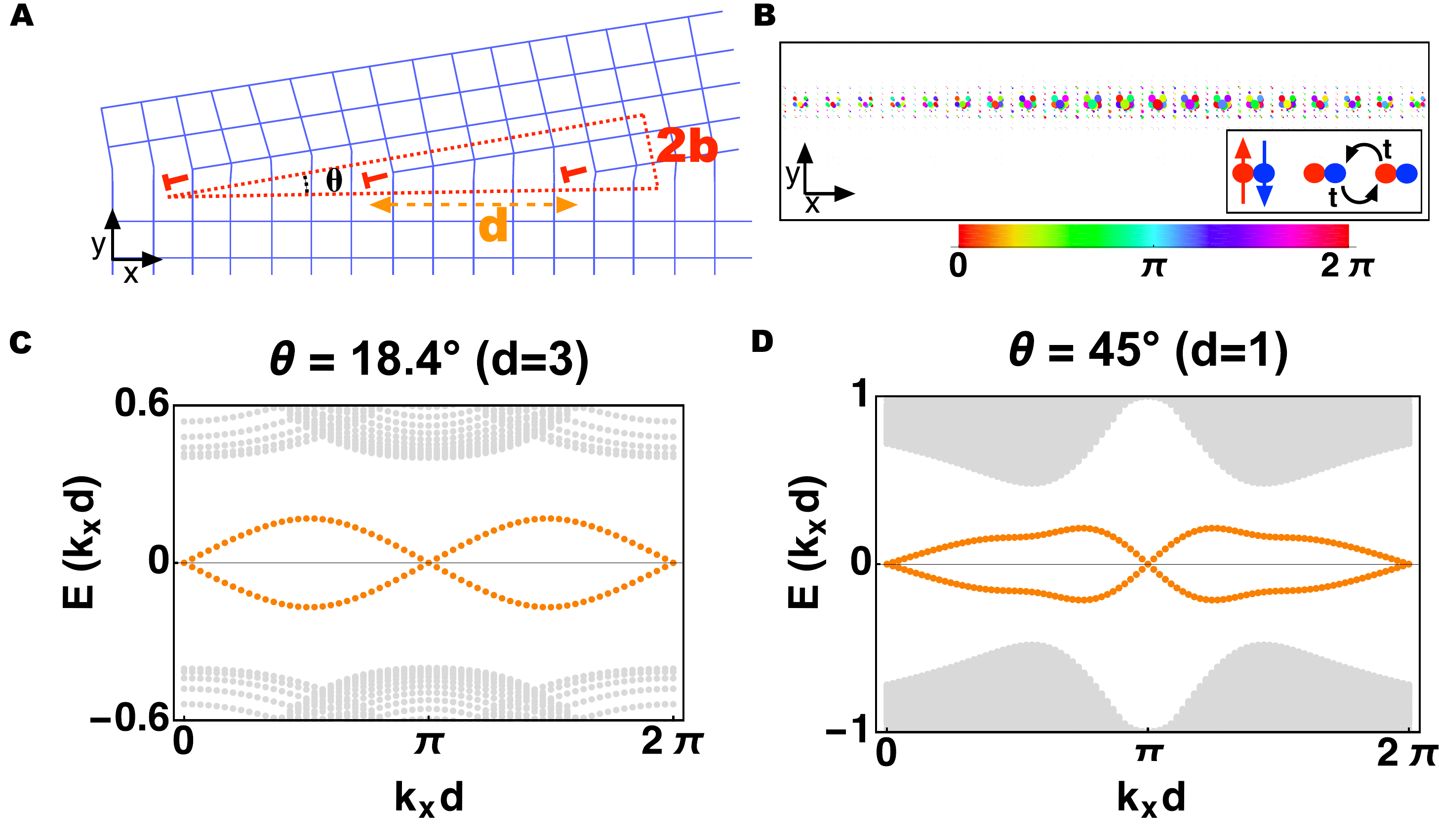}
\caption{\label{Fig::one}
Graphene-like semi-metal on a 1D grain boundary in a translationally active 2D TBI (the $M$-phase of the BHZ model on a square lattice, see Appendix \ref{App_BHZ}). (A) Schematic illustration of a GB. The coordination discrepancy due to the angular mismatch $\theta$ of lattice basis vectors results in an effective array of dislocations, each described by Burgers vector $\mathbf{b}$ and marked by a 'T' symbol, of spacing $d=|\mathbf{b}|/\tan \theta$. (B) Real space numerical tight-binding calculation of the low-energy states in the presence of a GB. The spinons bound to the dislocations hybridize with tunneling amplitude $t$, which gives rise to an extended 1D state along the GB. The radii of the circles indicate the amplitude of a wave function, associated with the node at $k_x=\pi/d$ of panel {\bf C}, while the colors indicate the phase. (C) The characteristic mid-gap bow-tie dispersion of the hybridized spinon bands (orange) and bulk bands (grey), corresponding to GB opening angle $\theta=18.4\degree$ for which isolated dislocations along the GB are well defined, as the function of the momentum $0 \leq k_x \leq 2\pi/d$ along the GB. (D) The same plot for the maximal opening angle $\theta=45\degree$. While the picture of isolated dislocations breaks down and the simple nearest-neighbour hybridization based sinusoidal dispersion is lost, the defining TRS semi-metal dispersion with the nodes at the TRS momenta survives. The data are for a $30\times60$ site or larger system with periodic boundary conditions.}
\end{figure*}

\section{Semi-metallic states on grain boundary lattice defects}

The physics of GBs can be derived from their elementary building blocks -- the lattice dislocations. It has been shown that when a 2D TBI is in a so called {\it translationally active topological phase} \cite{Slager13}, where the band topology is characterized by an odd number of band inversions at TRS momenta other than the $\Gamma$-point in the BZ, a lattice dislocation acts as an effective magnetic $\pi$-flux and binds a single Kramers pair of zero energy spinon modes localized at the core \cite{ Ran08, Qi08,Juricic12,Mesaros13}. These modes are 2D analogues of the famous spin-charge separated Jackiw-Rebbi soliton states \cite{Jackiw1976, Jackiw1981} realized in the Su-Schrieffer-Heeger (SSH) model of polyacetylene \cite{Su1980, Heeger1988}. The same mechanism applies also to 3D TBIs in translationally active phases, which generalize the characterization by weak indices \cite{Moore07,Slager13}. The natural lattice defect is a 1D dislocation line that behaves as $\pi$-flux tube and binds counter-propagating helical modes akin to edge states in 2D TBIs \cite{Ran09, Slager14}. A complete catalogue of modes bound to dislocations in generic 3D TBIs is given by the $K-b-t$-rule, that relates the spectrum of the surface states to the number of spinon modes bound to a dislocation line piercing the surface \cite{Slager14}.

When the spinon zero modes form an array, such as along a GB, one expects them to hybridize into an extended state. In the basis of two spinon modes, this state should be described by a two-band model $h(q)=h_0(q)+\mathbf{h}(q) \cdot \bm{\tilde{\sigma}}$. As $\bm{\tilde{\sigma}}$ act in the {\it spinon basis}, TRS requires $h_0(-q)=h_0(q)$ and $\mathbf{h}(-q)=-\mathbf{h}(q)$. Thus based on these very general considerations only, one expects a semi-metal with two TRS protected nodes at the TRS momenta $q=0$ and $q=\pi$. These nodes are degenerate when $h_0(q)$ either vanishes or is a constant, which, as we show below, is determined by the presence of bulk inversion symmetry. 

\begin{figure*}
\includegraphics[width=\textwidth]{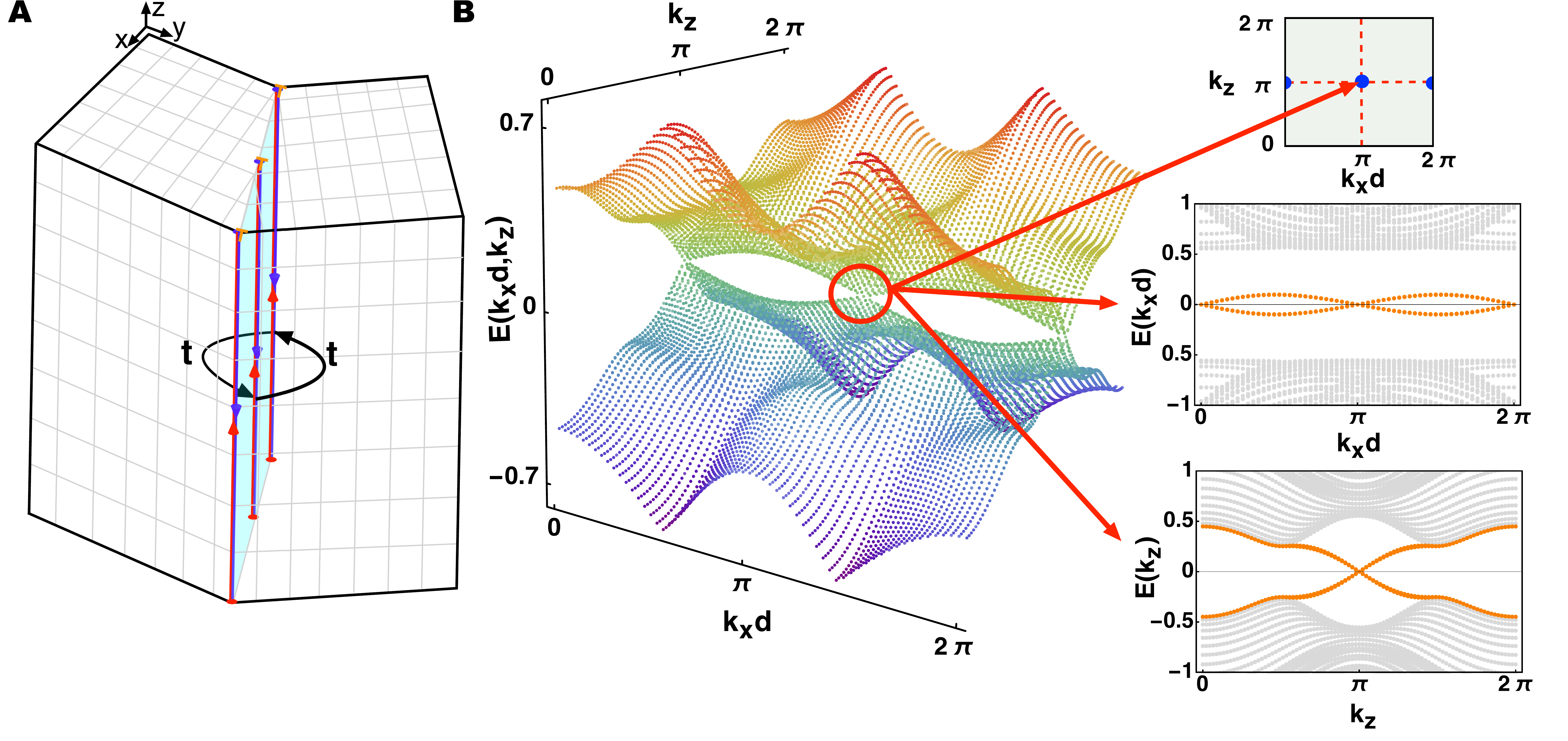}
\caption{\label{Fig::two}
Graphene-like semi-metal localized on a 2D grain boundary in a translationally active 3D topological band insulator (the $R$-phase of the BHZ model on a cubic lattice, see Appendix \ref{App_BHZ} for details). (A) Schematic illustration of the 2D GB, which now realizes a sheet of parallel 1D edge dislocation lines that extend along the grain boundary. Each edge dislocation hosts a pair of propagating helical modes localized along the dislocation cores that cross at $k_z=\pi$. (B) On a GB of opening angle $18.4\degree$, the hybridization of the helical modes results in the semi-metal band structure with two anisotropic pseudo-relativistic fermions at $(k_x,k_z)=(0,\pi)$ and $(\pi/d,\pi)$. Along the GB we recover the same mid-gap bow-tie dispersion as in the 2D case (shown top right for $k_z=\pi$), while along the edge dislocations the hybridized modes (orange) still flow into the bulk bands (gray) as a function of $k_z$ (bottom right for $k_x=\pi/d$). The data are for a $60\times60\times90$ system with periodic boundary conditions. }
\end{figure*}

To verify this prediction, we have carried out numerics for GB defects in the translationally active phases of BHZ tight-binding models for topological insulators with both TRS and inversion symmetry \cite{Bernevig06} (see Appendix \ref{App_BHZ} for details). Indeed, Figures \ref{Fig::one} and \ref{Fig::two} show that GBs support propagating states along their core. In the energy spectrum these states appear as emergent mid-gap 'bow-tie' bands that, reminiscent of the valleys in graphene, exhibit the expected two degenerate nodes at the two distinct TRS momenta. We have verified that these bands appear throughout the translationally active phases and persist for all GB angles including the maximal opening angle of $\theta=45\degree$ close to which a GB can no longer be approximated by an array of isolated dislocations. Thus when the chemical potential of the parent TBI is at the nodes, a self-organized graphene-like semi-metal emerges.

\section{Effective models for semi-metals on grain boundaries}

Having numerically established the existence of extended states on grain boundaries, we now turn to construct effective models for these semi-metals. The microscopic mechanism underlying their emergence is the hybridization of topological spinon modes bound to dislocations. We derive the hybridization induced dynamics by employing the construction for dislocation bound spinons that relies only on the existence of edge states originating from the band inversion momenta $\mathbf{K}_i$ and their symmetries \cite{Ran09}. The edge states relevant to construct the spinon modes can be identified using the $K-b-t$-rule \cite{Slager14}.

Consider a system of two coupled helical edges along $y$-axis described by the Hamiltonian 
\begin{equation}
H_0=v k_y \sigma^z \mu^z + m\mu^x,
\end{equation}
where $\sigma^z$ acts in the spin space, while $\mu^z=\pm 1$ denotes the two edges. In general, tunnelling of magnitude $m$ gaps out the edge states at the interface of the two TBIs. However, when an additional row of atoms for $y>0$ is inserted between the edges, or equivalently a dislocation with Burgers vector $\mathbf{b}=\mathbf{e}_x$ is created at $y=0$, the mass term for $y>0$ becomes $m e^{i\sum_i\mathbf{K}_i\cdot\mathbf{b}}$ \cite{Ran09,Slager14}. In a translationally active phase $\sum_i \mathbf{K}_i \cdot\mathbf{b}=\pi \ (\textrm{mod} \ 2\pi)$ and thus the sign of the edge state tunneling becomes $y$-dependent. The system is then equivalent to the 1D continuum SSH model with a mass domain wall \cite{Su1980, Heeger1988}. In a TRS system the domain wall binds a Kramers pair of localized spinons, which for the Hamiltonian $H_0$ live in a subspace described by the projector $P=(1+\sigma^z\mu^y)/2$.

To model a GB in a 2D translationally active topological material, we consider a dislocation array of spacing $d$ along $x$-axis. Employing the construction described above, the adjacent edge states then couple by a translationally and time-reversal invariant Hamiltonian 
\begin{equation}
H_{GB}=t(\cos k_x \mu^x + \sin k_x \mu^y),
\end{equation}
where $k_x$ spans the reduced BZ $0 \geq k_x \geq 2\pi/d$. The coupling strength $t$ follows from the overlap of the edge states, that is proportional to the overlap of the localized spinon wavefunctions (see Appendix \ref{App_spinons}), and thus scales as $t \sim d^{-1}$, implying that it is stronger for large opening angle GBs. Projecting this coupling into the spinon subspace, consistent with our prediction, we find that the spinons acquire dynamics described by
\begin{equation}
	P (H_{GB}) P = t\sin k_x \tilde{\sigma}^y,
\end{equation}
where $\tilde{\sigma}^y$ is an effective spin operator in the spinon subspace (see Appendix \ref{App_effective}). Comparing this form to our numerics, Figure \ref{Fig::one} shows that this simple expression indeed captures the defining nodal structure of the emergent mid-gap band for low opening angle GBs in the presence of spin-rotational and inversion symmetries (the large opening angle GBs require couplings beyond nearest neighbour).

In Appendix \ref{App_effective}, we have considered in detail the general case of breaking all the symmetries of the parent BHZ model. In the effective model $S_z$ conservation breaking via a Rashba term is modelled with $H_R=\alpha k_y \sigma^y$, inversion breaking with $H_I=\sum_i m_i \sigma^i \mu^y$ ($m_i$ can be either constant or proportional to $\cos k_x$) and TRS breaking via the Zeeman terms $H_B=\sum_i h_i \sigma^i$. In the presence of the Rashba term the projector to the spinon subspace is given by 
\begin{equation}
	P_\alpha= \frac{1}{2}\left[1+\frac{v\sigma^z+\alpha \sigma^y}{\sqrt{v^2+\alpha^2}} \mu^y\right]
\end{equation}	
and the effective Hamiltonian $H_{2D}=P_\alpha (H_{GB}+H_{I}+H_{B})P_\alpha$ becomes
\begin{equation}
	H_{2D} = \frac{m_y \alpha + m_z v}{\sqrt{v^2+\alpha^2}} + \left( \frac{vh_z + \alpha h_y}{\sqrt{v^2+\alpha^2}} + t\sin k_x \right) \tilde{\sigma}^y.
\end{equation}
This expression shows that the key property determining the response to TRS and inversion breaking is the spin texture of the edge states: Only perturbations  that couple to a spin orientation present in the edge states appear in the effective model (e.g., if $S_z$ is conserved, only perturbations proportional to $\sigma^z$ appear in the effective Hamiltonian). 

This effective picture is fully consistent with our numerics on the stability of the GB state (see Appendix \ref{App_numerics} for the details of the numerical calculations). First, $S_z$ conservation breaking Rashba coupling that is not strong enough to close the bulk band gap preserves the semi-metallic nodes. Second, terms breaking the bulk inversion symmetry in general shift the nodes to different energies, but, in contrast to interfaces of TBIs with different velocities \cite{Takahashi11}, can not gap them out. Third, TRS breaking terms gap the spectrum, but only if they do not anti-commute with $P_\alpha$ and only once their magnitude is comparable to the bandwidth $2t$ resulting from the hybridization. Furthermore, we have numerically found that no moderate random disorder can open up a gap at the nodes since the topological edge states underlying the hybridizing dislocation modes persist as long as the disorder does not drive the bulk out of the translationally active phase. This also implies that broken translational invariance along the GB, that can arise due to lattice reconstruction leading to disordered dislocation positions, does not pose a fundamental obstacle for the stability of the nodal band structure. Similar to chemical potential disorder that deforms dislocation mode wavefunctions, random bends along the GB result in random tunneling couplings $t$. In terms of the low-energy theory around the nodes, this gives rise to a random gauge field that only shifts the cones \cite{Foster06}. As long as this shift is smaller than the separation $\pi/d$ of the cones in the BZ, the semi-metallic behavior is stable. We have numerically verified this argument by considering random dislocation positions along the GB and found qualitatively similar stability to the case of chemical potential disorder. The self-organized semi-metal on the GB thus shares the topological stability of the edge states of the parent state: The nodes are degenerate in energy only if the bulk inversion symmetry is intact, but TRS is sufficient to protect the nodes themselves (no additional spatial symmetries with respect to the GB need to be assumed).

This mechanism generalizes straightforwardly to 3D translationally active TBIs, where a GB consists of a 2D sheet of parallel 1D dislocation lines, as illustrated in Fig.\ \ref{Fig::two}A. In a translationally active phase each dislocation binds a Kramers pair of helical modes \cite{Ran09,Slager14}. To derive a minimal effective model for their hybridization, we employ again the surface state construction with a coupling between adjacent surfaces. In the geometry of Fig.\ \ref{Fig::two}A, this is described by the Hamiltonian 
\begin{equation}
	H=v(k_y\sigma^y + k_z \sigma^z)\mu^z + m(y) \mu^x + H_{GB}.
\end{equation}
Similar to the 2D case, we project this Hamiltonian into the spinon subspace near $k_z=0$, which in the presence of TRS breaking Zeeman terms $H_B$ and inversion breaking terms $H_I$ gives the general minimal model (see Appendix \ref{App_effective})
\begin{equation}
	H_{3D} = m_y + v k_z \tilde{\sigma}^z +  \left(  h_y + t\sin k_x \right) \tilde{\sigma}^y.
	\label{Eq::Hameff_3d}
\end{equation}
This effective theory describes two linearly dispersing cones at the two TRS momenta $(k_x,k_z)=(0,\pi)$ and $(\pi/d,\pi)$ with anisotropic velocities $v$ along and $t$ perpendicular to the dislocation lines, respectively, consistent with Fig.\ \ref{Fig::two}B. This anisotropy is reduced for larger opening angles $\theta$ due to an enhanced overlap between the helical modes whose hybridization underlies the self-organized GB state. The two cones separated in the BZ are degenerate in energy if the bulk inversion symmetry with respect to the GB plane is preserved ($m_y=0$). On the other hand, if inversion is broken, the cones appear at different energies. As with the 2D case, we have numerically verified that the semi-metal is stable in the presence of moderate disorder and that the cones can only be gapped out with a TRS breaking Zeeman terms normal to the GB plane (see Appendix \ref{App_numerics}).

\section{Experimental consequences}

\begin{figure*}[ht]
\includegraphics[width=\textwidth]{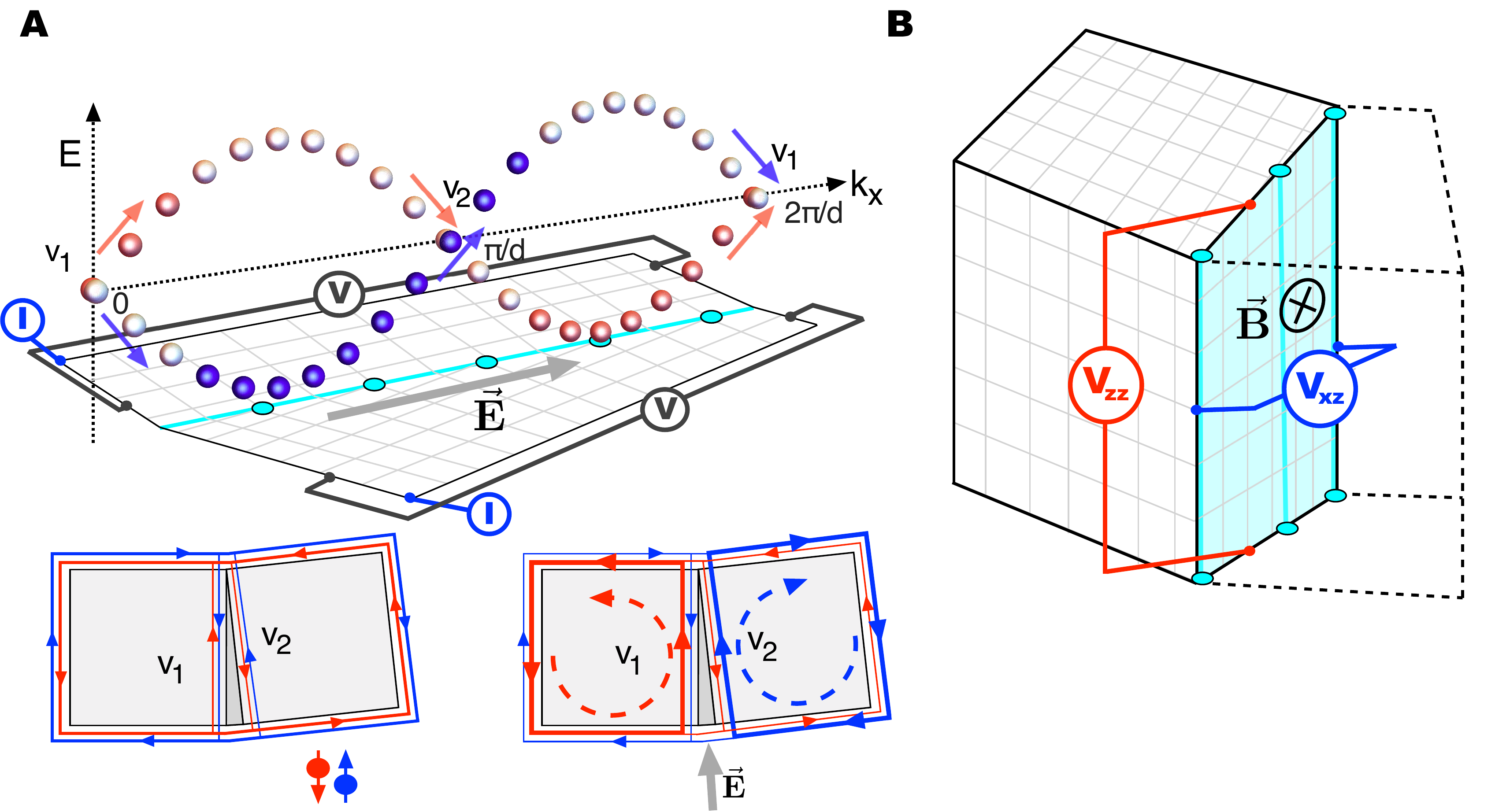}
\caption{\label{fig:exp} The experimental setups for identifying the signatures of the graphene-like semi-metals. (A) The one-dimensional bow-tie dispersion of the spinon semi-metal on a 1D GB implies a parity anomaly per spin component when an electric field $\vec{E}$ is applied along the grain boundary. The arrows indicate the shift in the spectrum from one valley to the other: When $S_z$ is conserved, valley $v_1$ accumulates an excess of spin down (red), while valley $v_2$ accumulates excess spin up (blue). These valleys can be associated with two co-existing channels that connect the helical edge states from the opposing surfaces. When $\vec{E}$ is applied along the GB, a current for both spin orientations is driven parallel to it. At the GB termination points this current is predicted to flow into the edge states that propagate to opposite directions resulting in a doubled spin Hall effect-like helical imbalance of the edge states on the two grains. Measuring the current imbalance $I$ is the hallmark signature of the valley anomaly exhibited by the spinon semi-metal. (B) The spinon semi-metal on a 2D GB may feature an odd-integer Hall effect with transverse conductivity $\sigma_{xz}=(2n+1)e^{2}/h$ in the presence of the perpendicular magnetic field $\vec{B}$. In the absence of external fields, another signature is provided by the diagonal ballistic optical conductivity of $\sigma_{zz}=(\pi/4) e^2/h$, which is half that of graphene. }
\end{figure*}

Having established the topological stability of the emergent semi-metals on GBs, we now turn to their experimental signatures. To detect the semi-metallic state on the 1D GB inside a 2D TBI, one can carry out a direct two terminal transport measurement analogous to the one used to detect edge states in a 2D quantum spin Hall insulator \cite{Koenig07}. When an electric field ${\bf E}$ is applied along the GB with leads attached to the GB ends, due to the two cones one should observe conductance of 
\begin{equation}
	\sigma=4e^2/h,
\end{equation}
i.e. twice the value measured for the QSH edge states. A more dramatic consequence of the helical bow-tie dispersion along a 1D GB is the existence of the parity anomaly \cite{Nielsen-Ninomiya1983} for each helical band. In the presence of two chiral cones, it gives rise to a {\it valley anomaly} that influences the edge transport. As illustrated in Fig.\ \ref{fig:exp}A, an electric field ${\bf E}$ along the GB generates excess helicity of opposite orientations at the two valleys that results in a net current flowing along the applied field. Associating the two valleys with two co-existing channels through which the helical edge states can flow from one GB termination surface to the other, the valley anomaly results in a helical imbalance of the edge currents at the two grains on each side of the GB.  This imbalance is proportional to the hybridization strength $t$ and it represents a hallmark transport signature of valley anomaly. To detect it, one can carry out a two-terminal edge transport measurement illustrated in Fig.\ \ref{fig:exp}A. When the two edges connected by the GB are biased by voltage $V$, a net current $I \sim t \frac{4e^2}{h} V$ is expected as a consequence of unequal helical currents. Recent experimental observation of semi-metallic transport on domain walls \cite{Ju2015, Ma2015}, such as in bi-layer graphene \cite{Martin2008, Zhang2013,Vaezi2013,jaskolski2016}, is highly encouraging that an experiment of this kind can also be carried out for GBs. In materials where spin is a good quantum number, i.e. spin-orbit coupling is negligible, the helical imbalance due to applied electric field translates into a spin imbalance of the edge currents on the two grains. In such cases the valley anomaly could also be detected by measuring the magnetic moment due to spin imbalance \cite{Nowack2013} or using Kerr rotation microscopy \cite{Kato2004, Sih2005}.

While the cones of the graphene-like semi-metal on a 2D GB can also be viewed as two co-existing channels through which the surface states of the 3D TBI can propagate between the surfaces connected by the GB, this state is experimentally most conveniently detected via two distinct "half-graphene"-like transport signatures. First, the measured ballistic optical conductivity is $\sigma_{zz}=(\pi/4) e^2/h$, which is half the value measured in graphene \cite{ballistic-minimal-conductivity}. Second, when a magnetic field is applied perpendicular to the GB in the setup shown in Fig.\ \ref{fig:exp}B, the non-degeneracy of each cone implies a contribution of $(n+1/2) e^2/h$ to the Hall conductivity, which in turn implies odd-integer quantum Hall effect with total Hall conductivity  
\begin{equation}
\sigma_{xz}=(2n+1) e^2/h,
\end{equation}
when the cones are degenerate in energy (bulk inversion symmetry is present). As there is always a known contribution from the surface states, it is in principle possible to extract the signal associated with the Hall conductance arising solely from the GB semi-metal. The emergent chiral symmetry of the effective model (\ref{Eq::Hameff_3d}) also suggests the existence of edge states on the 1D GB edges on the 2D surfaces (see Appendix \ref{App_effective}). These can contribute additional Fermi arc-like features to the surface states of the parent 3D TBI, which may be detectable via zero frequency optical conductivity measurements \cite{Grushin15}, or give rise to GB edge transport when the cones occur at different energies due to broken inversion symmetry \cite{Ramamurthy15}.

\section{Discussion and Outlook}

We have shown that TRS protected semi-metals -- helical wires exhibiting a valley anomaly in 2D and an anisotropic graphene-like state showing an odd-integer quantum Hall effect in 3D -- can emerge on grain boundaries in TBIs. These states are self-organized and enjoy the same topological stability as the TBI edge states. The only requirement for their emergence is the TBI to be of a translationally active type where lattice dislocations bind a Kramers pair of localized modes. As grain boundaries occur naturally in crystalline samples, or they can be experimentally manufactured \cite{Wang11}, the challenge is to identify translationally active materials. A convenient guide to candidates is given by the space group classification of TBIs \cite{Slager13}. For modelling purposes we used the $M$- and $R$-phases of the 2D square and 3D cubic lattice, respectively, with the latter having a potential realization in electron-doped BaBiO$_2$ \cite{Yan2013}. Nonetheless, our mechanism is completely general and the outlined GB physics should therefore also appear in the already experimentally verified translationally active materials. This rapidly expanding list includes Bi$_x$Sb$_{1-x}$ with band inversions at the $L$-points \cite{Hsieh2008}, the topological Kondo insulator SmB$_6$ with band inversions at three $X$ points \cite{Xu2014, SmB6} and the recently discovered bismuth iodide compounds that feature a transitionally active phase with band inversions at the $Y$ points \cite{Autes2015}. Topological semi-metals are expected also on GBs in the topological crystalline Sn-based compounds  \cite{Hsieh12, Dziawa2012, Tanaka2012}, that feature band inversions at the $L$ points as long as the protecting symmetry is respected \cite{Slager13, Slager14}, as well as in Bismuth bilayers with a band inversion at the $M$ point, where a GB state has been identified in a recent {\it ab initio} study \cite{DFT-Nunes}.

Grain boundaries provide a natural setting for arrays of localized topological solitons to hybridize in any crystalline materials. The emergent extended states depend on the topology and the symmetry class of the parent state. Our general mechanism is readily applicable to model the distinct GB states that may be realized in topological states of matter with symmetries different than considered here. For instance, lattice dislocations in Sr$_2$RuO$_4$, a candidate for chiral $p$-wave superconductor, can host Majorana modes \cite{Hughes14}, which are expected to hybridize into novel superconducting states \cite{Lahtinen12}. A full classification of extended states on GBs in crystalline topological states of matter provides for a fascinating subject of future work. Finally, our results demonstrate the potential of extended lattice defects in exploring manifestations of electronic band topology beyond by now well understood surface states.

\section*{Acknowledgements}

This work was supported by the Dutch Foundation for Fundamental Research on Matter (FOM).  V.\ J.\ acknowledges the support of the Netherlands Organization for Scientific Research (NWO). V.\ L.\ acknowledges the support from
the Dahlem Research School POINT fellowship program.

\clearpage
\newpage

\appendix

\section{Tight-binding BHZ models for numerics }
\label{App_BHZ}

\subsection{The 2D BHZ model}

The 2D BHZ model is defined on a square lattice \cite{Bernevig06} with two spin degenerate $|s\rangle$ and $|p_{x}+ip_{y}\rangle$ type orbitals on every latttice site. In natural units $\hbar=c=e=1$, the model is defined by the nearest neighbour tight-binding Hamiltonian
\begin{equation}\label{Hamiltonian::TB}
\mathcal{H}_{\text{TB}}=\sum_{\mathbf{r}, \boldsymbol{\delta}} (\Psi^{\dagger}_{\mathbf{r}} T_{\boldsymbol{\delta}}\Psi_{\mathbf{r}+\boldsymbol{\delta}}+\text{H.c.}) +\sum_{\mathbf{r}}  \Psi^{\dagger}_{\mathbf{r}} \mu \Psi_{\mathbf{r}}.
\end{equation}
Here  $\{\boldsymbol{\delta}\}=\{ \mathbf{e}_x,\mathbf{e}_y \}$ denote the vectors connecting the nearest neighbor sites and $\Psi^\top_{\mathbf{r}}=(s_{\uparrow}(\mathbf{r}),p_{\uparrow}(\mathbf{r}), s_{\downarrow}(\mathbf{r}), p_{\downarrow}(\mathbf{r}))$ annihilates the $s$ and $p$
type orbitals at site $\mathbf{r}$. The tunneling of the spin up and spin down electrons is given by
\begin{equation*}
T^{*}_{\boldsymbol{\delta}, \downarrow \downarrow}=T_{\boldsymbol{\delta}, \uparrow \uparrow}=
\begin{pmatrix}
\Delta_{s}& t_{\boldsymbol{\delta}}/2\\
t'_{\boldsymbol{\delta}}/2&\Delta_{p}
\end{pmatrix},
\end{equation*}
that describes inter-orbital tunneling  $t_{\boldsymbol{\delta}}=-i\exp({i\varphi_{\boldsymbol{\delta}}})$ and  $t'_{\boldsymbol{\delta}}=-i\exp({-i\varphi_{\boldsymbol{\delta}}})$ whose phase is given by the polar angle $\varphi_{\boldsymbol{\delta}}$ of the vector $\boldsymbol{\delta}$, and intra-orbital tunneling of magnitude $\Delta_{s/p}=\pm B$.
The on-site energies are parametrized as $\mu=(M-4B)\tau_{z}\otimes\sigma_{0}$, where the Pauli matrices ${\boldsymbol \tau}$ and ${\boldsymbol\sigma}$ act in the orbital and spin space, respectively. Here, ${\tau}_0$ and ${\sigma}_0$ are the $2\times2$ identity matrices. By performing a Fourier transform, the Hamiltonian \eqref{Hamiltonian::TB} assumes the block-diagonal form
\begin{equation}\label{Hamiltonian::TriRec}
\mathcal{H}_{\text{TB}}=\sum_{{\bf k}}\Psi_{\bf k}^\dagger
\begin{pmatrix}
H(\mathbf{k})&0\\
0&H^{*}(-\mathbf{k})
\end{pmatrix}\Psi_{\bf k}.
\end{equation}
The Hamiltonians for each spin component can be decomposed as $H(\mathbf{k})={\boldsymbol\tau}\cdot{\bf d}({\bf k})$, where the vector ${\bf d}(\mathbf{k})$ has the components $d_{x,y}(\mathbf{k})=\pm\sin(k_{x,y})$ and $d_{z}=M-2B(2-\cos(k_{x})-\cos(k_{y}))$. The spin up and down blocks are related by the time-reversal symmetry represented by an antiunitary operator ${\cal T}=\tau_0\otimes i\sigma_y K$, with $K$ as the complex conjugation. The energy spectrum for each spin component is given by $E(\mathbf{k})=\pm|{\bf d}|$, while the spectrum of the full Hamiltonian is doubly degenerate.

The dispersion $E(\mathbf{k})$ is gapped except for the values $M/B= 0, 4$ or $8$, where the gap closes at the $\Gamma$ $(0,0)$, $X$ $(\pi,0)$ and $Y$ $(0,\pi)$, or $M$ $(\pi,\pi)$ points of the Brillouin zone, respectively. When $0<M/B<4$ the system is in a topological $\Gamma$-phase, while for $4<M/B<8$ the system is in a topological $M$-phase. For other values of $M/B$ the system is topologically trivial and does not have helical edge states. The $\Gamma$- and $M$-phases are characterized by the same $Z_2$ invariant and both exhibit helical edge states. They are distinguished by lattice defects that break translational symmetry: In the $\Gamma$-phase a lattice dislocation does not bind localized modes, while in the $M$-phase they act as $\pi$-fluxes and bind localized zero energy modes \cite{Juricic12}. Hence, we refer to the $M$-phase as being {\it translationally active}. In this phase the edge states appear always at the projection of the $M$-point to the surface BZ, i.e. they cross at $k=\pi$ on the edge BZ.

The 2D BHZ model has additional symmetries besides TRS. Explicitly, all the symmetries comprise:
\begin{itemize}
\item Time-reversal symmetry $T H(k) T^{-1}=H(-k)$ represented by $T=i(\sigma_{y}\otimes \tau_0) K$ satisfying $T^2=-1$, where and $K$ denotes complex conjugation.
\item Particle-hole symmetry $P H(k) P^{-1}=-H(-k)$ represented by $P=(\sigma_0 \otimes \tau_x) K$ satisfying $P^2=1$.
\item Chiral symmetry $C H(k) C^{-1}=-H(k)$ represented by $C=PT=i\sigma_{y} \otimes \tau_x$.
\item Spin-rotation symmetry $S_z H(k) S_z^{-1}=H(k)$ represented by $S_z=\sigma_z \otimes \tau_0$.
\item $C_4$ crystalline point group symmetry in the x-y-plane represented by $R=\frac{1}{\sqrt{2}}(e^{i\frac{\pi}{4}}+e^{-i\frac{\pi}{4}} \tau^z)e^{i\frac{\pi}{4}\sigma^z}$ that maps $k_{x,y} \to \pm k_{x,y}$, $\tau^{x,y} \to \mp \tau^{y,x}$ and $\sigma^{x,y} \to \pm \sigma^{y,x}$. It gives rise to an inversion symmetry $I H(k) I^{-1}=H(-k)$ satisfying $I^2=-1$ represented by $I=R^2=i\sigma^z\otimes\tau_{z}$, which can be further decomposed into inversions along x- and y-axes as $I=I_x I_y$, where $I_x=i\sigma^x \tau^0$ and $I_y=i\sigma^y\tau^z$ obey $I_i H(k_i) I_i^{-1}=H(-k_i)$.
\end{itemize}
The $S_z$ symmetry implies that the model can be viewed as two copies of opposite chirality Chern insulators, one for each spin component. This conservation is non-generic and is in general broken by spin-orbit coupling. For the BHZ model specific to HgTe quantum wells, the orbital dependent Rashba form reads
\be
	T_{\boldsymbol{\delta}} \to T_{\boldsymbol{\delta}} + i \frac{R}{2} (\tau_0 + \tau_z) \mathbf{e}_z \cdot (\boldsymbol{\sigma} \times \boldsymbol{\delta}).
\ee
This term breaks also the accidental $P$ and $C$ symmetries, leaving the model with only TRS and the $C_4$ point group symmetry. The latter can be further broken, while preserving TRS, by terms of the form $H_I = \sum_i m_i \sigma^i \tau^i$, with non-zero $m_i$ breaking the inversions $I_i$. The coefficients $m_i$ are constant if they originate from local potentials, or proportional to $\cos k_i$ when they result from tunneling processes.

\subsection{The 3D BHZ model}

The 2D BHZ model on a square lattice can be generalized to a 3D cubic lattice. The key difference is that in 3D the model can no longer be viewed as two decoupled Chern insulators of opposite chirality, and an $S_z$ conservation breaking spin-orbit coupling is needed for the model to exhibit topological phases. Working directly in the momentum space, the tight-binding model contains two spin degenerate orbitals, which are in general described by a Hamiltonian of the form
\begin{equation*}\label{eq::general}
H=\epsilon(\mathbf{k})\mathbf{1}+\sum_{\alpha}{d}_{\alpha}(\mathbf{k})\gamma_{\alpha}+\sum_{\alpha\beta}{d}_{\alpha\beta}\gamma_{\alpha\beta},
\end{equation*}
where $\gamma_{\alpha}$ are the five Dirac matrices obeying the Clifford algebra $\{\gamma_\alpha,\gamma_\beta\}=2\delta_{\alpha\beta}$, $\gamma_{\alpha\beta}$ are the ten commutators $\gamma_{\alpha\beta}=\frac{1}{2i}[\gamma_{\alpha},\gamma_{\beta}]$.
Specifically, we take the following representation for the $\gamma$-matrices
\begin{align}
\gamma_{0} & =\sigma_{0}\otimes\tau_{z}\qquad \gamma_{1}=\sigma_{x}\otimes\tau_{x}, \qquad \gamma_{2}=\sigma_{y}\otimes\tau_{x}, \\ \qquad\gamma_{3} \nonumber & =\sigma_{z}\otimes\tau_{x}, \qquad \gamma_{5}\equiv - \gamma_0\gamma_1\gamma_2\gamma_3=\sigma_{0}\otimes\tau_{y}, \nonumber
\end{align}
with ${\bm \sigma}$ and ${\bm \tau}$ being again the standard Pauli matrices acting in the spin and orbital space, respectively. Taking into account only nearest-neighbor terms and setting the lattice constant to identity, the BHZ model in 3D is described by the Hamiltonian \rf{eq::general} where $d_{\alpha\beta}=0$, $d_0 = M-2B(3-\cos k_{x} -\cos k_{y} -\cos k_{z} )$ and $d_{1,2,3} = \sin k_{x,y,z}$. As with the 2D model, $B$ is again the magnitude of the intra-orbital hopping and $M$ is the difference of the onsite energies between the two orbitals. However, one should keep in mind that while this Hamiltonian bears formal similarity to the 2D one \rf{Hamiltonian::TriRec}, the explicit spin-orbit coupling is included in the $\gamma$-matrices that describe spin-flipping inter-orbital tunneling, whose magnitude has been set to unity. The phase diagram exhibits now topological phases for $0 < M/B < 4$, $4 < M/B < 8$ and $8 < M/B < 12$. Only the last, the $R$-phase with band inversion only at the $R$-point $(\pi,\pi,\pi)$, is a translationally active phase where the surface states always cross at $(\pi,\pi)$ on the surface BZs \cite{Slager13}.

This model has also additional symmetries originating from the orbital space. In addition to TRS, the model has the global particle-hole and chiral symmetries described by $P=K\sigma^y\tau^y$ and $C=i\tau^y$, respectively. The cubic lattice point group symmetry gives rise to inversion $I=\tau^z$, which can be again decomposed into inversion $I_i=i\sigma^i\tau^z$ normal to each of the three axes. The $P$ and $C$ can again be broken by $H_I=\sum_i m_i \sigma^i \tau^y$ without breaking TRS, with each non-zero $m_i$ breaking again also the inversion $I_i$. Even in the presence of such terms, the 3D model still has a residual particle-hole symmetry $\tau^x H(\mathbf{k})\tau^x=-H(-\mathbf{k})$, which can be further broken by $m_0 \tau^x$ that also breaks all inversion symmetries.

\subsection{Grain boundaries in tight-binding models}

\begin{figure}[ht]
\includegraphics[width=0.3\textwidth]{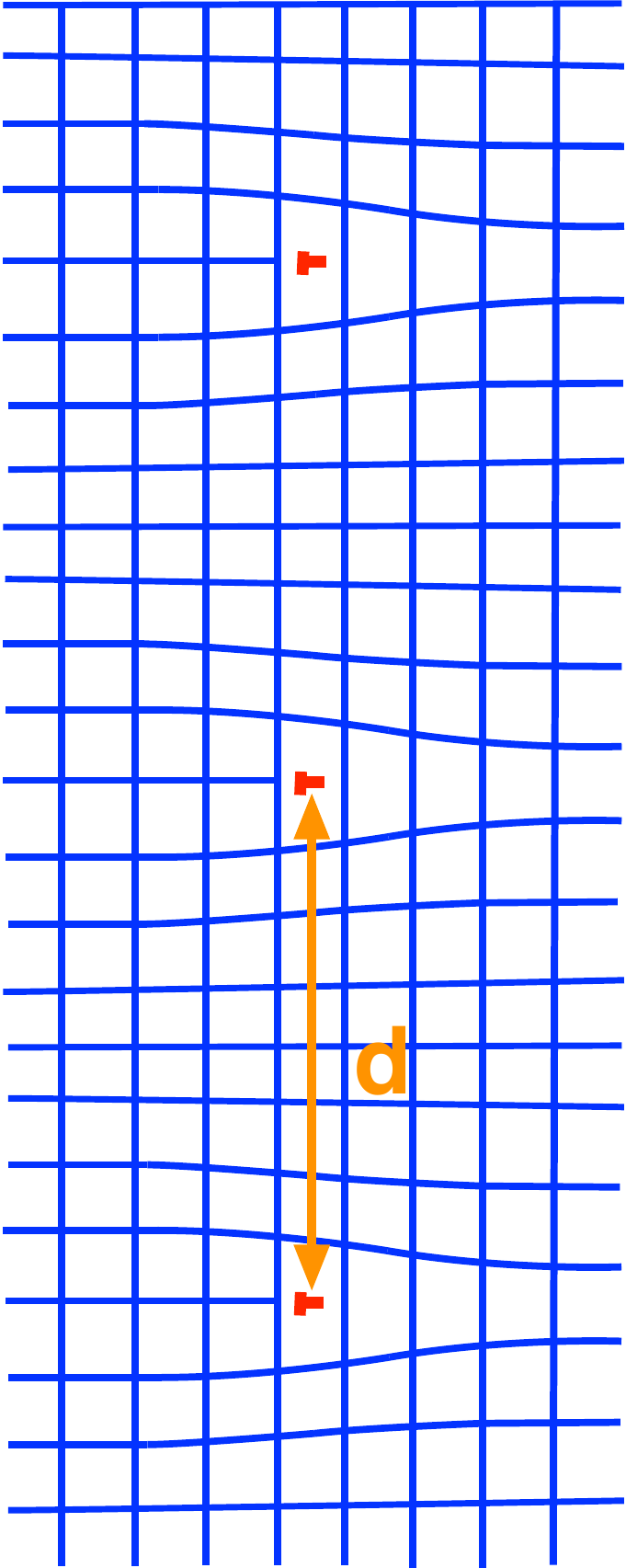}
\caption{\label{Fig:lattices} Illustration of the grain boundary lattice used in the simulations in the particular instance of spacing $d=8$. }
\end{figure}

To simulate a GB in the tight-binding numerics, we consider a lattice that consists of two parts with different number of rows glued together across the GB defect. A lattice with a GB of spacing $d$ with translational invariance in $y$-direction is created by taking the two halves with $L_x$ and $L_x /(d+1)$ rows of sites and leaving every $(d+1)$th row unconnected, a representative of which is illustrated in Fig. \ref{Fig:lattices}. Finally, we corroborated our results with an alternative procedure of simply matching two perfect crystalline regions under an angle, schematically illustrated in Fig.\ 1A of the main text. The resulting system with coordinate discrepancy then produced the same qualitative results, as illustrated with the $45\degree$ result presented in the main text.

\section{Derivation of effective models for the GB states}
\label{App_effective}

\subsection{An effective model for the 1D GB inside a 2D TBI}

We find the effective theory for the GB by coupling together adjacent the helical edge states of QSH insulator with a periodic array of trenches with dislocation lines inserted, following the construction  of Ref \cite{Ran09} for solitons bound to dislocations. In the absence of Rashba coupling $S_z$ is conserved and two decoupled helical edge states along $y$-axis are described by the Hamiltonian $H_0 = v  k_y \sigma^z \mu^z$, where the $\sigma$-matrices act in the spin space and the $\mu^z=\pm 1$ denotes the two edges. Proximity tunneling between the edge states is described by the term $m\mu^x$, which in general gaps the edge states and merges the two QSHE insulators into a single one. When a semi-infinite row of atoms is inserted for $y>0$, creating a dislocation of Burgers vector $\mathbf{b}=e_x$ at $y=0$, the tunneling term giving mass to the edge states across the inserted row of atoms becomes $m \to e^{i\sum_i \mathbf{K}_i \cdot \mathbf{b}}$ \cite{Ran09,Slager14}, where $\mathbf{K}_i$ are the band inversion momenta from which the edge states originate. In other words, in a translationally active phase where $\sum_i \mathbf{b} \cdot \mathbf{K}_i=\pi \ (\textrm{mod} \ 2\pi)$, the mass term becomes position dependent such that $m(y<0)>0$, while $m(y>0)<0$. The low-energy theory for a single trench is then given by
\begin{equation}
	H_0 = v  k_y \sigma^z \mu^z + m(y) \mu^x,
\end{equation}
which is equivalent to a continuum theory for a SSH model with mass changing sign at $y=0$. Such model is known to have two bound state solutions at the mass domain wall, one for each spin component. The general form of these localized solutions is given by $\psi = e^{\frac{1}{v} \int m(x) dx} \phi$, with the four-spinor obeying $\sigma^z \mu^y \phi = \phi$. This condition defines the subspace where the localized soliton solutions live, with the projector onto this subspace  $P=(1+\sigma^z \mu^y)/2$. Considering a translationally invariant array of such bound modes, their hybridization can be modelled by coupling the edge states from adjacent trenches by introducing
\begin{equation}
	H_{GB} = t \left( \cos(q) \mu^x + \sin(q)\mu^y \right).
\end{equation}
Projecting this down to the spinon subspace yields the minimal effective model $P H_{GB} P = t \sin(q) \tilde{\sigma}^y$, where the effective spin operator is defined by $\tilde{\sigma}^y = \mu^y$.

The effective model $H_0+H_{GB}$ has additional symmetries represented by  $S_z=\sigma^z$, $P=K\sigma^z\mu^z$, $C=\sigma^x\mu^z$ and $I=i\sigma^x\mu^x$ symmetries besides TRS. To derive an effective model for the fully symmetry broken case, we consider edge states with generic spin texture $H_0 = k (\mathbf{v} \cdot \bm{\sigma}) \mu^z + m(y) \mu^x$, where we have defined the normalized vector $\mathbf{v}=(v_x,v_y,v_z)/|\mathbf{v}|$. The spinons live then in a subspace defined by the projector $P=[1+(\mathbf{v} \cdot \bm{\sigma})\mu^y]/2$, which breaks the $S_z$, $P$ and $C$ symmetries. To consider the effect of breaking the inversion symmetry and TRS, we introduce the perturbations $H_I=\sum_i m_i \sigma^i \mu^y$ and $H_T=\sum_i h_i \sigma^i$, respectively. Here the magnitudes $h_i$ of the Zeeman terms are constants, while in general the inversion-symmetry breaking arises from edge state tunneling and thus $m_i \sim \cos(q)$. Projecting all these down to the spinon subspace gives the effective model $H_{2D}=P(H_{GB}+H_I+H_T)P$, or explicitly
\begin{align}
	 H_{2D} =  \mathbf{m}\cdot \mathbf{v} + \left(t\sin q + \mathbf{h}\cdot \mathbf{v} \right)\tilde{\sigma}^y.
\end{align}
The spectrum is given by $E_\pm = \mathbf{m}\cdot \mathbf{v} \pm \left(t\sin q + \mathbf{h}\cdot \mathbf{v} \right)$, which for $\mathbf{v}=(0,\alpha,v)/\sqrt{\alpha^2+v^2}$ reduces to the one presented in the main text. The key observation is that the effect of the (spin-rotational symmetry breaking) spin-orbit coupling is to change the spin texture of the edge states, and that controls which terms perturb the effective model. Only inversion or TRS breaking perturbations with a component parallel to the spin texture appear in the effective model, which explains why in our numerics only certain perturbations affect the spectrum in the full 2D BHZ model. In general, inversion breaking terms shift the nodes at the TRS momenta to different energies, while TRS breaking terms gap them out once their magnitude is of the order of the bandwidth $t$ resulting from the hybridization. Other TRS invariant tunneling terms can also be added to the Hamiltonian, but we have checked that they only deform the band structure, while preserving the key features summarized above. In the numerics on the BHZ model, they are responsible for the suppression of the bandwidth of the semi-metal when spin-rotational symmetry is broken.

\subsection{Effective model for the 2D GB inside a 3D TBI}

Similar analysis can be carried out to derive an effective minimal model for the 2D GB. Let us assume that the dislocation sheets are inserted in $y-z$-plane, such that the dislocation lines run in $z$-direction, while the GB is oriented in $x$-direction. The coupled surface states in the presence of the dislocation sheet are then described by the Hamiltonian
\begin{equation}
	H_0 = v  (k_z \sigma^z + k_y \sigma^y) \mu^z + m(y) \mu^x,
\end{equation}
which is a direct generalization of the 2D case. $k_z$ is still a good quantum number and soliton solutions at the dislocation can be found immediately considering the $k_z=0$ case. The projector onto their subspace is now given by $P=(1+\sigma^y \mu^y)/2$.

Allowing again for a generic spin texture along the hybridization direction and including also the inversion and and TRS breaking terms, in direct analogy to the 2D case, we arrive at the effective model
\begin{equation}
	H_{3D} = \mathbf{m} \cdot \mathbf{v} + v k_z \tilde{\sigma}^z + (\mathbf{h} \cdot \mathbf{v} + t\sin q)\tilde{\sigma}^y,
\end{equation}
where we have defined the effective spin operators $\tilde{\sigma}^z=\sigma^z\mu^z$ and $\tilde{\sigma}^y=\mu^y$. For $\mathbf{v}=(0,1,0)$ this reduces again to the case presented in the main text. As with the 2D case, we find that the key property determining the response to symmetry breaking perturbations is the spin texture of the surface states.

\subsubsection{Fermi arc-like surface states on GB edges}

In the presence of the inversion symmmetry ($\mathbf{m}\cdot \mathbf{v}$=0), the effective model possesses also chiral symmetry. This means that it can be viewed as a 1D SSH model \cite{Su1980}, but with a GB momentum $q$ dependent mass $M(q)=\mathbf{h} \cdot \mathbf{v} + t\sin q$. Due to the chiral symmetry, this model has two topologically distinct phases: for $M(q)<0$ the band structure is topologically trivial and characterized by a winding number $\nu=0$, while for $M(q)>0$ the winding number evaluates to $\nu=\pm 1$ and the system is in a topological phase and exhibits edge states. Thus when the TRS breaking is negligible compared to the hybridization, the mass depends on the momentum $q$ along the GB such that $M(q)>0$ for $0 < q < \pi$, while $M(q)<0$ for $\pi < q < 2\pi$, and the edge states of the GB semi-metal span only half of the 1D edge Brillouin zone of the GB.

Thus when the 2D GB inside a 3D TBI terminates on the 2D surface, we predict the extended GB state to contribute additional Fermi arc-like features on the surface states. These states appear as localized flat bands at the GB edges parallel to the Burgers vector describing the dislocations that constitute the GB. Since in a real material the GB edge BZ spans only the reduced BZ $0 < q < 2\pi/d$, these features are expected to be more pronounced for GB defects with a large opening angle. As long as they do not hybridize with the TBI surface states, which does not happen as long as the surface cones occur at different momenta, they can co-exist and could in principle be detected via low-frequency optical conductivity measurements \cite{Grushin15}. Alternatively, breaking inversion symmetry will in general shift the cones to different energies, which makes the flat edge states dispersing and can lead to edge currents at the GB edges \cite{Ramamurthy15}.

\section{1D tight-binding picture for spinon hybridization}
\label{App_spinons}

Here we present a complementery picture in terms of localized spinon wavefunction hybridization. It enables us to directly connect the spinon hopping magnitude $t$ to the hybridization energy $\epsilon$ that arises when two spinon binding dislocations are in proximity of each other.

In the context of the 2D BHZ model, the localized spinon zero modes $\Phi_s=\sum_r \phi_{s}(\mathbf{r}) \Psi_{s,\mathbf{r}}$ can be solved analytically for a vanishing Rashba coupling near $M/B \to 8$ \cite{Juricic12,Mesaros13}. For single spin component with the dislocation located at the origin, the modes are given by the orbital space spinors
\be \label{WF}
  \phi_\uparrow ({\bf r}) = \frac{ \sin(\lambda r) }{N \sqrt{r}} e^{-r/\xi} \left( \begin{array}{c} e^{-i\theta} \\ i \end{array}\right),
\ee
where we use polar coordinates $\mathbf{r} =(r,\theta)$, $\xi = 2B$ is the correlation length, $\lambda = \sqrt{1-4MB}/2B$ and $N$ is a normalisation constant. TRS requires that $\phi_\downarrow ({\bf r})=\phi_\uparrow^* ({\bf r})$ and guarantees that even if these two modes are localised on the same dislocation, they do not hybridize even in the presence of a Rashba coupling (although the wavefunctions and the orientation of the spinors in the combined spin-orbital space will be modified). Thus isolated dislocations deep in the $M$-phase bind a Kramers pair of exponentially localized modes, whose wavefunctions exhibit spatial oscillations at the wavelength $\lambda$, which depends on the microscopic parameters.

\begin{figure*}[ht]
\includegraphics[width=\textwidth]{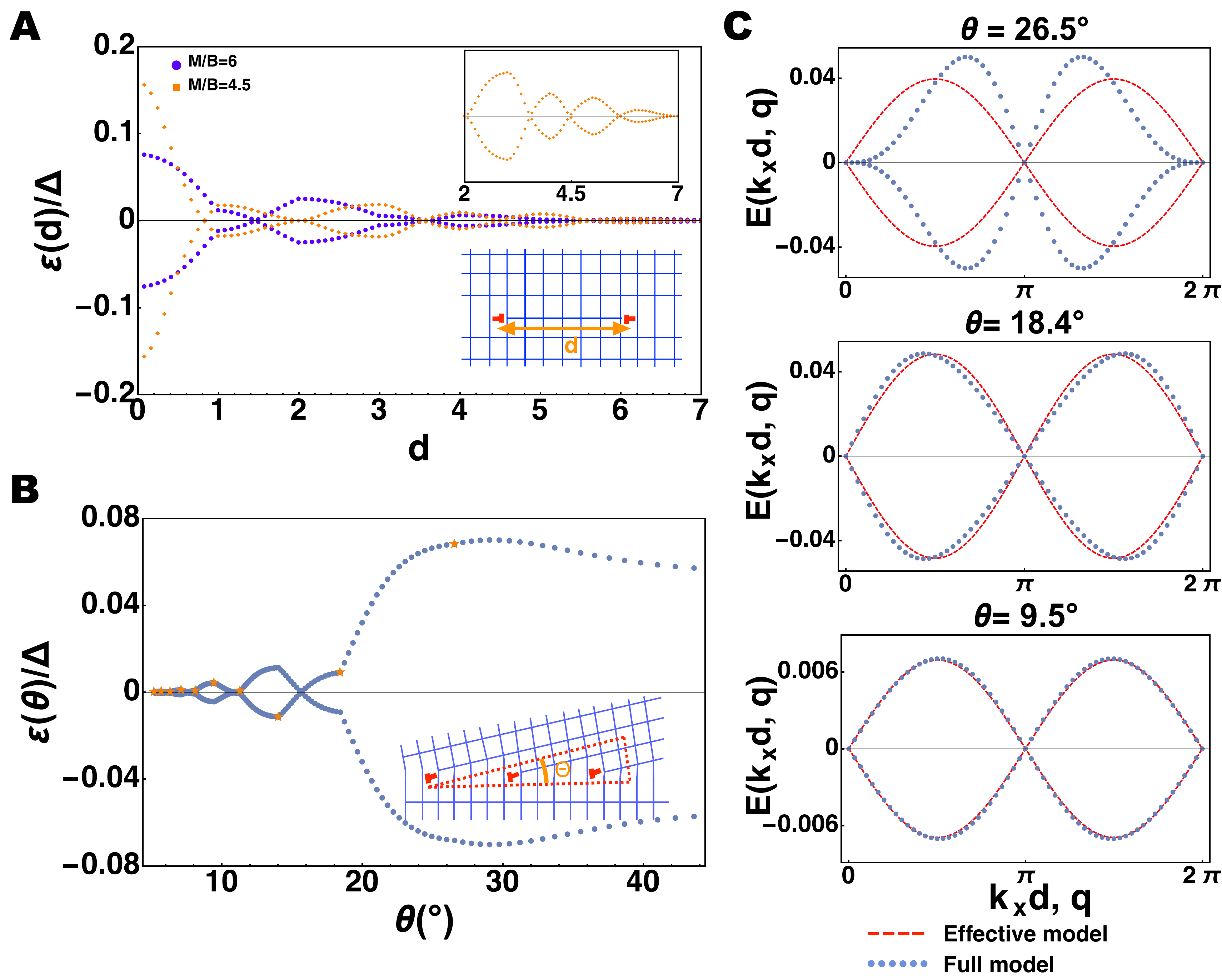}
\caption{\label{Fig:hybr} {\bf A} The hybridization energy $\epsilon$ as the function of the separation $d$ between two dislocations in units of the bulk band gap $\Delta$. For different values of $M/B$ the hybridization energy always exhibits exponential decay and characteristic oscillations that exhibit clear sinusoidal oscillations at larger separations (inset). The kinks as functions of $d$ are artefacts of simulating smooth dislocation transport by adiabatically switching on and off the relevant couplings. {\bf B} The two dislocation hybridization energy $\epsilon$ in the GB geometry as the function of the GB angle $\theta$ in units of the bulk band gap $\Delta$. {\bf C} The mid-gap spectra of the 2D BHZ model across the reduced Brillouin zone $0 < k_x <2\pi/d$ for different GB angles $\theta=\textrm{arctan}(|b|/d)$ and the approximation of each by the unperturbed minimal effective model $H_{2D}=t\sin q$ over $0<q<2\pi$ with the hybridization energy $\epsilon(d)=t/2$ as the only input. As the dislocation separation $d \sim \theta^{-1}$ increases, the nearest-neighbour approximation becomes more valid and we find excellent agreement. All data has been produced using a 90 $\times$ 60 system with $M/B=6$.}
\end{figure*}

In the presence of many dislocations, all the modes $\Phi_{i,s}$ bound to each dislocation remain near zero energy as long as the separation $d$ between all the dislocations satisfies $d \gg \xi$. When two dislocations are brought into proximity ($d \approx \xi$), the wavefunctions overlap. This gives rise to a finite (possibly spin-flipping) tunneling amplitudes $t_{ss',|i-j|} \sim \int \textrm{d} {\bf r} \phi_{s,i}^\dagger ({\bf r}) \phi_{s',j} ({\bf r})$. This hybridizes the spinon modes and gives rise to a pair of Kramers degenerate modes at finite energy $\epsilon \sim t$ (assuming $S_z$ conservation, i.e. $t=t_{\uparrow\uparrow}=t_{\downarrow\downarrow}$ and $t_{\uparrow\downarrow}=t_{\downarrow\uparrow}=0$), localized simultaneously on both dislocations. In principle, the hybridization energy $\epsilon$ could be analytically obtained, but as the analytic solution is available only in the $M/B \to 8$ regime, we perform instead a numerical analysis. Figure \ref{Fig:hybr}A shows that the two-dislocation hybridization energy obeys the qualitative form $\epsilon(d) \propto \cos(\lambda d) e^{-d/\xi}$ throughout the $M$-phase, with the wavelength $\lambda$ decreasing with smaller $M/B$. These oscillations in the hybridization energy, that can cause change the sign as the function of dislocation separation $d$ and thus energetically favour different specific separations, follow directly from the oscillating tails of \rf{WF} and they are qualitatively similar to $\pi$-flux vortices in other Dirac-like systems \cite{Cheng09, Lahtinen11}. We also find that while the localized spinon modes \rf{WF} possess full rotational symmetry near $M/B \to 8$, for smaller values they become anisotropic as the band gap minimum switches from the $M$-point (full rotation symmetry) to the $X$- and $Y$-points ($C_4$ symmetry). It follows that also the hybridization energy becomes anisotropic depending on the microscopic parameters. This may have consequences on the nature of the emergent state if the several GBs formed arrays of same dimensionality as the parent state.

A grain boundary at small opening angles $\theta$ is well approximated by a 1D array of lattice dislocations. These again hybridize pairwise, with the hybridization energy $\epsilon$ as the function of the GB opening angle $\theta$ shown in Figure \ref{Fig:hybr}B. It is thus natural to ask whether the bandwidth of the 1D GB can be traced back to the pairwise hybridization energy $\epsilon$? To study this, we take $t=2\epsilon$ as the simplest ansatz. This is motivated by the fact that TRS and translationally-invariant spinful free 1D fermions with imaginary tunneling amplitude $i\epsilon$ have the spectrum $E_\pm = \pm 2\epsilon \sin q$. Figure \ref{Fig:hybr}C shows that as the opening angle $\theta$ of the grain boundary gets smaller, i.e. the spacing $d=|\mathbf{b}|/\tan\theta$ between the dislocations becomes larger and nearest-neighbour approximation is justified, the spectrum of the mig-gap GB bands in the full BHZ model is indeed accurately approximated with the relevant two-spinon hybridization energy $\epsilon(d)$ as the only input. This directly connects the more general edge state construction to the hybridization of the localized spinon modes and corroborates this picture in the context of the BHZ models.

\section{Disorder analysis of the grain boundary states}
\label{App_numerics}

Here we present numerical data on the stability of the graphene-like spinon semi-metal when random chemical potential disorder is included in the parent TBI. The GB states are expected to be stable for two physically motivated complementary reasons: (i) The spinons are rooted in the edge states, which can not be removed by moderate disorder and (ii) the localized spinon wavefunctions are only locally deformed, which can only result in random tunneling along the GB, but not in a random chemical potential that can gap the modes out. From the point of view of the spinon tight-binding model the latter translates to a random gauge field around the chiral cones that only shifts them around, but can not gap them out unless the mean displacement is of the order of the cone separation in momentum space. Thus large opening angle GBs are expected to be most stable.

\begin{figure*}[ht]
\includegraphics[width=\textwidth]{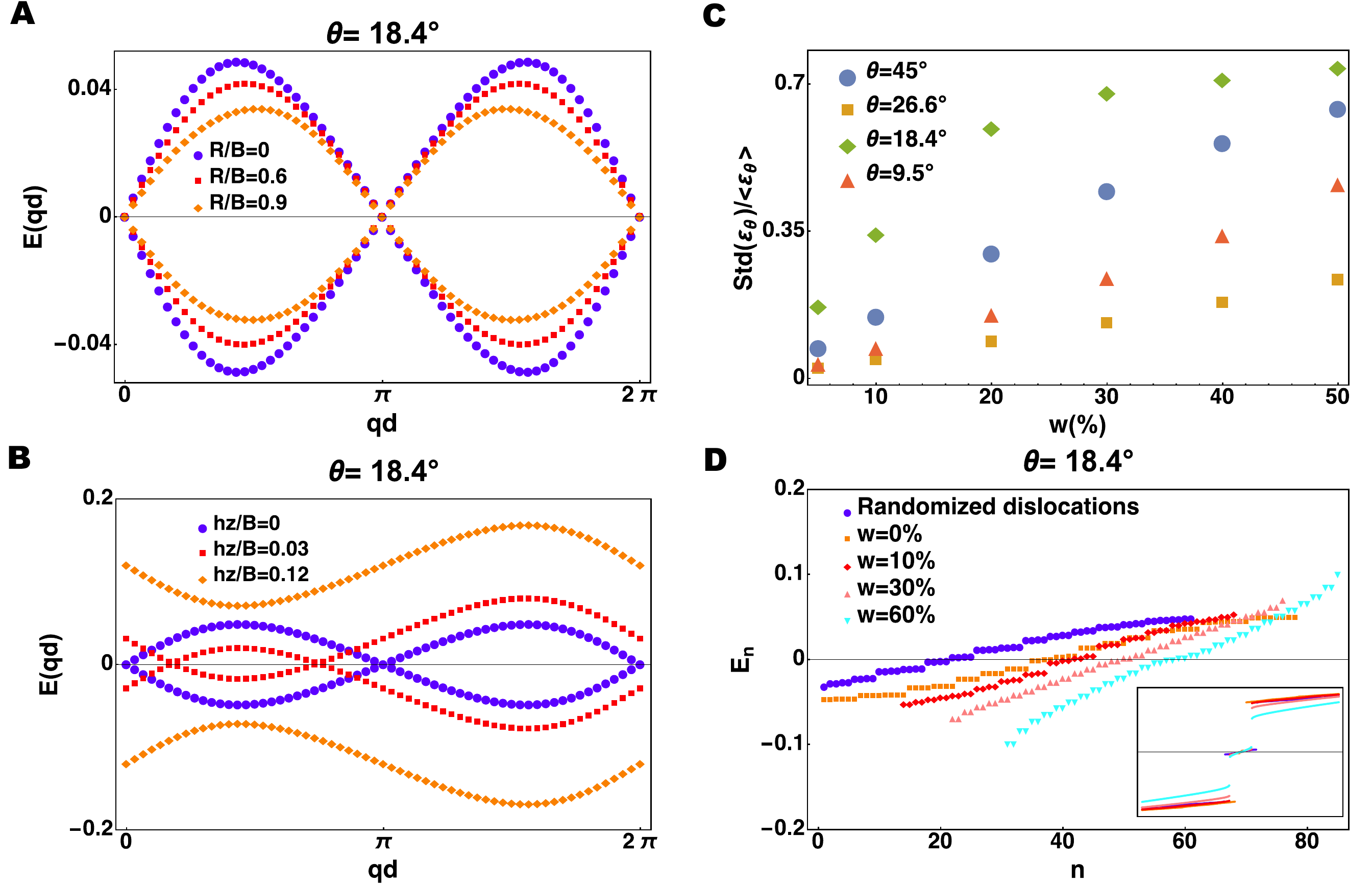}
\caption{\label{Fig::SOM2} Stability of the graphene-like spinon semi-metal. {\bf A} Inclusion of a Rashba term of magnitude $R$ breaks $S_z$ conservation, which leads to suppression of the bandwidth and different velocities at the cones, but TRS guarantees that the cones are stable.  {\bf B} An $S_z$ conservation preserving Zeeman term of magnitude $h_z$, gives rise to a mass for the spinons. In the hybridized spectrum it shifts the spinon bands to opposite directions in energy and thereby moves the cones towards each other. The cones annihilate and gap out at a quadratic band touching point when $h_z=2\epsilon$. {\bf C} When local chemical potential disorder of magnitude $w$ is introduced, i.e. $\mu$ is a random variable picked from box distribution $[(1-w)(M-4B), (1+w)(M-4B)]$, the hybridization energies $\epsilon$ become also disordered with the fluctuations, as quantified by the standard deviation Std$(\epsilon)$ around the mean $\langle \epsilon \rangle$ increasing monotonously. The rate of increase depends on the microscopic details around the nearest-neighbour two dislocation hybridization energy corresponding to different GB angles $\theta$ as show in Figure \ref{Fig:hybr}B. {\bf D}  In the full 2D TBI, we find that the gapless spinon semi-metal persists at least up to 50$\%$ chemical potential disorder ($n$ enumerates the states), which is insufficient to close the bulk gap (inset). We have also simulated randomized dislocation positions (dislocations along a GB of spacing $d$ are randomly displaced by up to $d$ lattice constants) and find similar stability. All data has been produced using a 90 $\times$ 60 system in 2D with $M/B=6$. The disorder data has been produced for system size 42 $\times$ 30 with periodic boundary conditions and averaged over 100 disorder realizations.}
\end{figure*}

To substantiate these general arguments, we have studied the effect of local random disorder on the bi-partite hybridization energies $\epsilon$ that enter the 1D tight-binding model as tunneling amplitudes. Figure \ref{Fig::SOM2}C shows data for the 2D BHZ model demonstrating that even if the disorder averaged hybridization energies are strongly influenced by local random disorder, disorder strengths of up to 50$\%$ are insufficient to cause random sign flips. Thus while any disorder potential contains a Fourier component that corresponds to nodal scattering (perfect sign dimerization would gap out nodes even at maximal separation of $\pi$), we expect the spinon semi-metals to be very stable. Indeed, numerical data presented in Figure \ref{Fig::SOM2}D shows remarkable resilience with gapless mid-gap bands again persisting for disorder strengths of up to 50$\%$. Due to the general arguments given above, similar stability applies also to the 2D GB semi-metal in the 3D BHZ model.

This stability analysis also implies that perfect translational invariance of the GB is neither a necessary condition for the stability of the GB semi-metals. Random spinon tunneling can arise also due to random turns along the GB and thus the stability under local random disorder also implies stability in the absence of disorder, but with broken translational invariance along the GB. Indeed, Figure \ref{Fig::SOM2}D shows that the gapless mid-gap state persists also for randomized GB positions along the GB. We therefore conclude that translational invariance on average is sufficient for graphene-like semi-metal to emerge along them.

Finally, in Figure \ref{Fig::SOM2}B we have also considered effects of TRS breaking Zeeman terms. In agreement with the effective minimal model derived from the edge states, we find $h_z$ shifting the $S_z$ conserving spectrum in energy and gapping it when $h_z > 2t$. No other Zeeman term orientation affects the spectrum as the edge spins are polarized in $z$-direction. When Rashba term is included and the edge states pick up a component parallel to the propagation in $y$-direction, we find also $h_y$ having similar effect. Likewise, we have checked the same edge state spin texture dependence of the inversion breaking terms that, consistent again with the effective model, in general shift the nodes to different energies, but do not gap them out.

\end{document}